\documentclass[nonacm]{acmart}
\makeatletter
\def\@ACM@checkaffil{
    \if@ACM@instpresent\else
    \ClassWarningNoLine{\@classname}{No institution present for an affiliation}%
    \fi
    \if@ACM@citypresent\else
    \ClassWarningNoLine{\@classname}{No city present for an affiliation}%
    \fi
    \if@ACM@countrypresent\else
        \ClassWarningNoLine{\@classname}{No country present for an affiliation}%
    \fi
}
\makeatother
\usepackage{xcolor}
\usepackage{graphicx}
\usepackage{wrapfig}

\begin{document}
\title{On the Political Economy of Link-based Web Search}

\setcopyright{none}

\author{Deepak P\hspace{0.02in}}
\affiliation{Queen's University Belfast, UK}
\email{deepaksp@acm.org}
\author{James Steinhoff\hspace{0.02in}}
\affiliation{University College Dublin, Ireland}
\email{james.steinhoff@ucd.ie}
\author{Stanley Simoes\hspace{0.02in}}
\affiliation{Queen's University Belfast, UK}
\email{ssimoes01@qub.ac.uk}

%
%
%
%

%
\begin{abstract}
Web search engines arguably form the most popular data-driven systems in contemporary society. They wield a considerable power by functioning as gatekeepers of the Web, with most user journeys on the Web beginning with them. Starting from the late 1990s, search engines have been dominated by the paradigm of link-based web search. In this paper, we critically analyze the political economy of the paradigm of link-based web search, drawing upon insights and methodologies from critical political economy. We draw several insights on how link-based web search has led to phenomena that favor capital through long-term structural changes on the Web, and how it has led to accentuating unpaid digital labor and ecologically unsustainable practices, among several others. We show how contemporary observations on the degrading quality of link-based web search can be traced back to the internal contradictions with the paradigm, and how such socio-technical phenomena may lead to a disutility of the link-based web search model. Our contribution is primarily on enhancing the understanding of the political economy of link-based web search, and laying bare the phenomena at work, and implicitly catalyze the search for alternative models. 
\end{abstract}

\maketitle              
\makeatletter \gdef\@ACM@checkaffil{} \makeatother


\section{Introduction}

The World Wide Web, conceived as a 'universal linked information system' and 'a place to be found for any information or reference which one felt was important'~\cite{berners1989information}, was heralded as an enabler for democratization of information. On the eve of its thirtieth anniversary in 2019, its founder, Tim Berners-Lee, released an open letter~\cite{tim201930} outlining the {\it 'sources of dysfunction'} affecting the Web, critiquing its evolution over the three decades. Among those critical references is that it encompasses {\it 'system design that creates perverse incentives where user value is sacrificed'}. In this context, it is interesting to see that companies with a significant interest in the web have occupied four spots among eight in the list of trillion-dollar companies\footnote{\url{https://en.wikipedia.org/wiki/List\_of\_public\_corporations\_by\_market\_capitalization}}. What led to the evolution of the Web from being an information seeking platform, to being what increasingly looks like a business platform? Towards understanding this, we start from the 'gatekeepers'\footnote{\url{https://digital-markets-act.ec.europa.eu/index\_en}} of the Web, the search engines~\cite{germano2020opinion}, where most people start their Web journeys; search engines probably form the most popular information retrieval systems that have ever existed. In particular, we consider the political economy of the dominant paradigm of web search, {\it link-based web search}. We take a critical political economy approach towards understanding link-based web search, locating our work within critical algorithm studies. 

We start with outlining several possible conceptualizations of web search, illustrating how the contemporary link-based paradigm of search is only one among a wide variety of available choices. From this vantage point, we briefly consider the first paradigm of modern web search, content-based search, and its various aspects and vulnerabilities. Against this historical backdrop, we look at its successor, the paradigm of link-based search, the dominant paradigm of web search since the early 2000s, which is also arguably the only existing model in today's web. We first look at the technology of link-based search, and analyze the choice architecture embedded within it from technological, philosophical and political perspectives. We outline the variety of critical observations of link-based search from a wide spectrum of scientific scholarship. We motivate the need for an understanding of the political economy of link-based search that focuses on the totality of the socio-technical ecosystem and social relations embedded within it. As a step towards such a goal, we outline a theoretical framework that focuses on the variety of consequences of link-based search. We describe the various first order and second order consequences within our framework, tracing how these relate to the variety of critical observations outlined earlier. We also consider how these effects synergize to reduce the utility of link-based search, and also locate link-based search within the broader context of automation within critical political studies. 

Our analyses of the political economy of link-based search lead us to conclude that there is an intrinsic alignment between libertarian political ethos and the socio-technical ecosystem of link-based search. Consequently, web search engines may be functioning as a substantive pathway for capital (as understood within Marxist studies~\cite{marx1867capital}) to operate on the Web. The intent of our work is to contribute to the scholarship in critical algorithm studies and aid to advance awareness of the politics of web search to levels where imagining responsible and fairer alternatives becomes possible. 



\section{Web Search: Conceptualization, Socio-technical Challenges and Practice}

We start with possibilities of conceptualizing web search, and the various challenges and possibilities that exist therein. From this vantage point, we will also consider content-based search, the precursor of the link-based model. 


\subsection{Conceptualizing Web Search}\label{ref:conceptual}

If we position web search as a gatekeeper of the Web, where the Web is envisioned to be {\it 'a place to be found for any information or reference which one felt was important'}~\cite{berners1989information}, how should web search function? 

Towards conceptualizing a possibility, we consider the notion of {\it use value} from critical political studies. To quote from {\it Capital}~\cite{marx1867capital}, {\it 'the utility of a thing makes it a use-value'}. Elsewhere, Marx~\cite{marx1859contribution} says: {\it 'A use value has value only in use, and is realized only in the process of consumption.'} Use value, in Marxist literature, is often contrasted with exchange value, the worth of the thing expressed in monetary terms on the market. 

From this perspective, web search may be seen as a marketplace that brokers the matching between {\it information needs} expressed as web search queries and {\it use values} that manifest as web pages.  Use values are not absolute but relative to the need; thus, each page may be regarded as having a use value distribution over the space of needs. For example, a government page with immigration information may have a high use value for a user seeking to apply for a visa, whereas it may not have any use value who is searching for football scores of a recent match. When some web pages are returned by a search engine in response to a search query and the user clicks on one, the user can be regarded to have {\it consumed} the use value represented by the web page that has been clicked on. This consumption also leads to a realization of the use value of the web page. Thus, the {\it visibility} of a web page through a search engine directly relates to realization of use values embedded within it. There are at least two models to think of, while designing web search from a clean slate. We look at each one in turn. 


\subsubsection{Use-value Oriented Search} This model asserts that web pages be catered to search queries in such a way that the {\it visibility} is proportional to {\it use values} conditioned on the information need; we summarize this model as {\bf be true to use values}. In other words, across user queries associated with a particular information need, two web pages will get visibility in proportion to their use values for the information need. This is a conceptual model which doesn't prescribe what ought to be retrieved for {\it each} query, but merely asserts that web pages be accorded visibility, {\it across queries}, in proportion to their use value. Thus, this model is a web page creator oriented perspective. If visibility through a search engine functions as an incentive to drive the creation of web pages, this model would power the creation of more and more use values on the web, potentially bringing it closer to Berners-Lee's idealism of the Web as {\it 'a place to be found for any information or reference which one felt was important'}.

\subsubsection{Need Oriented Search} The need oriented model simply suggests that search engines should focus on satisfying needs, which could be stated as {\bf provide the best use values for the needs expressed in the query}. This model implicitly ensures customer satisfaction in that each user is provided with the most relevant web page for her need. This, as maybe obvious to the reader, is the predominant conceptual model that contemporary web search engines adhere to. The need-based model could interpret a query on the basis of the user profile or previous searches; thus, personalization and 'query-less' search~\cite{yin2022exploiting} also often falls under the need based model. 


\subsubsection{Use-value Orientation vis-a-vis Need Orientation} There are potentially several factors that lead to the pervasiveness of the need oriented model, and the virtual absence of the use-value oriented model among contemporary search engines. {\it First}, web search is naturally viewed as a process triggered by the submission of the search query, making the need-oriented approach much easier to conceptualize technically. {\it Second}, with most web search engines being profit-seeking commercial entities that compete with one another, the apparent alignment of the need-oriented model towards market values such as customer satisfaction leads to a preference towards the need oriented model in the contemporary neoliberal world order. {\it Third}, web pages are not exhausted on user consumption, thus can be recommended any number of times. This makes need oriented operation technically simple. {\it Fourth}, web pages are often perceived as passive entities as opposed to representing active users to which the web search engine is accountable to. This makes the web page creator focused perspective embedded in the use value oriented model less prevalent. In contrast to the web search case, elements of a use value oriented model has been explored in retrieval/recommendation systems where accommodation options~\cite{biega2018equity}, job seekers~\cite{singh2018fairness,zehlike2017fa}, profiles of convicted criminals~\cite{zehlike2017fa} and e-commerce products~\cite{patro2020fairrec} are sought to be retrieved. In such recommenders, it is important to keep the human actors behind the retrieved entities in good humor or else they may withdraw from the system. In contrast, it is assumed that web pages pose no such threat to search engines. Yet, as we will see, actions by human actors behind web pages, i.e., webmasters, could have significant consequences to search engines. Following similar lines, a recent article~\cite{sundin2022whose} points out the need to consider search engines as multi-sided platforms rather than a user-focused single-sided platform. 


\subsubsection{Realizing Need Oriented Search} 

While the targeted need oriented model is conceptually simple, there are significant socio-technical impediments to realizing it in practice. We outline some main considerations herein vis-a-vis practices in contemporary search technologies. 

\noindent{\bf Query to Need Mapping:} From the early history of search engines, it has been observed that users tend to prefer short queries. Among the first quantitative studies on query logs~\cite{silverstein1999analysis} (for the AltaVista search engine) in 1999 indicates that user queries contain 2.35 words on an average. Similar statistics were observed for the Excite search engine in 2001~\cite{spink2001searching} where the average query was found to be 2.4 words. More recent studies have recorded an increase in average query length, with a 2009 study~\cite{zhang2009time} reporting 2.9 and a 2012 study~\cite{taghavi2012analysis} recording 3.08. Yet, it is apparent that these remain significantly shorter than the size of verbal communications between humans while conveying information needs. The short query lengths pose a significant challenge to map queries to {\it information needs} that they relate to. While this has been a subject of research (e.g.,~\cite{figueroa2015exploring}), there are potential intrinsic ambiguities. As authors imply in a recent paper~\cite{haider2022algorithmically}, a query containing the names of two cities could correspond to a variety of information needs viz. identifying travel options, comparing cost-of-living or liveability, or other kinds of needs. The social preference for short search queries - which may arguably be accentuated with the prevalence of simple search interfaces promoted by search engines - predicates a deep technical challenge of mapping queries to needs. Note that it may be impossible to accomplish this mapping by any amount of sophisticated technology since the translation of the information need in the mind of the user to a representation in language may be intrinsically understood as a lossy process. As the noted philosopher Wittgenstein argues~\cite{Wittgenstein_notebook}, the fact that a language needs to have certain properties makes it impossible to convey {\it everything} that is intended (p108). 

\noindent{\bf Estimating Use Value:} The estimation of the use value of a web page to a particular information need is another facet of socio-technical complexity. As pointed out earlier, the use value of a web page needs to be assessed in relation to the information need. Use value is often interpreted as the notion of {\it relevance}, one that the Cambridge dictionary defines as: {\it 'the degree to which something is related or useful to what is happening or being talked about'}. The assessment of relevance is confounded by the fact that the query-to-need mapping is unlikely to be done unambiguously. This potentially means that relevance may need to be assessed over a distribution of needs than a sole need. Further, there exist many candidate notions of relevance, as outlined recently~\cite{sundin2022whose}. While systemic relevance focuses on the match between the query and the web page (thus, correlating to the idea of use value), user relevance focuses on how well the individual users may subjectively evaluate the web page. A third notion of societal relevance may require that web pages be also judged on beneficence to the society, thus excluding potentially pernicious web pages (e.g., sexist and racist ones~\cite{noble2018algorithms}). Of the several alternatives available, contemporary search engines arguably use only a narrow definition of relevance. In particular, most search engines use an interpretation that stretches the notion of user relevance towards maximizing engagement; notably, engagement could be interpreted as a query-agnostic notion. For example, prior research from Microsoft~\cite{song2013evaluating} has explored user engagement retention in the face of relevance drop (such as in data voids~\cite{mager2023advancing}). 

\noindent{\bf Presenting Search Results:} Search engines have, since early days, uniformly adopted the one-dimensional presentation model of search results as a list. This is typically accomplished as an ordering of search results in the descending order of relevance, arguably optimized for the user propensity to scan results from top to bottom~\cite{chierichetti2011optimizing}. This contrasts with presentation of e-commerce and image search results in a two-dimensional matrix form. User perusal of results have been found to follow an F-shape pattern~\cite{nielsen2006f}, which may, in simple terms, be understood as the attention being more dense within a triangle with its corners in the top-left, top-right and bottom-right of the display (the {\it golden triangle}~\cite{chierichetti2011optimizing}). The rapid drop in the quantum of attention as one moves down the list, called {\it position bias}, has led to a practical difficulty in ensuring that user attention is apportioned to web pages based on their relevance, resulting in a stream of work on addressing the disparity between user attention and relevance~\cite{zehlike2022fairness}. 

\begin{wrapfigure}[10]{r}{0.4\textwidth}
  \begin{center}
    \includegraphics[width=0.38\textwidth]{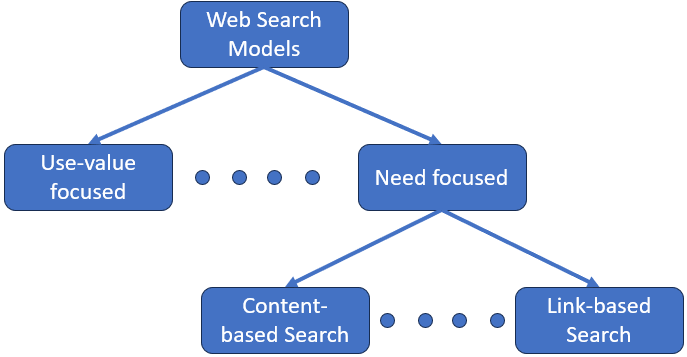}
    \end{center}
  \caption{Possible Conceptualizations of Web Search}
  \label{fig:conceptualizations}
\end{wrapfigure}

\subsubsection{Summary} We have seen in this section that there are at least two distinct models of conceptualizing web search, and these could have sub-models, as indicated in Fig~\ref{fig:conceptualizations}. Of these, we observe that the need oriented model has virtually been the only model that's used now.  Our mention of use value based search is to indicate a potential alternative conceptualization, and the remainder of the text will focus on need-based search. We also saw how need based web search can be operationalized through decoupled tasks, a mapping from queries to needs, and an estimation of need-based use values. The socio-technical impediments, engendered by a mix of human preferences and technical complexity, make it challenging to realize need-based search. On the other hand, such socio-technical challenges create a matrix of possibilities to approximate need based search by choosing different trade-offs and design choices.

\subsection{Content-based Web Search}

While the focus of this paper is link-based web search, we glance quickly at content-based web search, its precursor, for a historical context. Early web search engines such as AltaVista\footnote{https://en.wikipedia.org/wiki/AltaVista}, the major pre-Google search engine launched in 1995, prided primarily on their comprehensiveness of coverage and expressivity of search\footnote{Refer AltaVista firsts at: https://digital.com/altavista/}. While search algorithms tend to be proprietary, their workings can be inferred based on the nature of popular search engine optimization (SEO) tactics targeting them. The predominance of {\it keyword stuffing} and {\it hidden text} as major SEO strategies of the 1990s\footnote{https://www.techradar.com/news/black-hat-seo-7-tactics-to-leave-back-in-the-90s} indicate that pre-Google search engines such as AltaVista and Yahoo! were largely based on text/content based search. 

The content-based paradigm would determine the inclusion of web pages in search results based on the extent of the match between keywords in the query and those in the web page. While the precise technical details may vary - some keywords may be considered as more important than other, or other variants - a web page would have a higher chance of figuring in results associated with search queries built out of words that appear in its content. This paradigm notably short-circuits the identification of the {\it need}, and yields a technologically simple operation. 

While content-based search may have been a reasonable approximation of need based search, the emergence of SEO as a dominant force alters the landscape. With the pervasiveness of the Web and the pre-eminence of search engines as an arbiter of user attention on the Web, the determination of which products, services and messages get noticed is increasingly taken up by search engines~\cite{lewandowski2023jasist}. This creates incentives for webmasters\footnote{Webmasters are the creators/managers of web pages.} to operate in ways that align with search engine algorithms. In particular, in attempting to enhance the visibility of their website over content-based search engines, they would need to use {\it abundance of keywords aligned with user queries for which the web page would be relevant}, in order to ensure visibility. Such practices would invalidate the effectiveness of content overlap metric as an approximation of need-based search. 

\subsection{Vulnerability to Goodhart Effects}

It is well understood in literature that {\it "any observed statistical regularity will tend to collapse once pressure is placed upon it for control purposes"}. This is often referred to as the Goodhart's law~\cite{goodhart1984monetary}. In the case of content-based search, the usage of content-match as a proxy for relevance functions as a {\it perverse incentive} for web page creators to optimize for content match leading to a divergence between content match and intended relevance. Such effects are also variously referred to as Campbell's law~\cite{campbell1979assessing} and {\it Cobra effect}~\cite{siebert2001kobra}. The latter refers to anecdotal experiences in colonial India where bounties were offered for dead cobras in order to stem the menace caused by the snake. As the bounty got entrenched in society, people apparently began breeding cobras for the income, leading to an increase in the menace rather than a decrease. Content-based search, due to having a simple structure, could be gamed {\it at the level of individual webmasters} using methods such as keyword stuffing as seen earlier. The emergence of SEO services at the time also indicate the entrenchment of perverse incentives. In particular, this is an instance of adversarial Goodhart within the four categories of Goodhart effects~\cite{manheim2018categorizing}. While content-based search may be regarded as particularly vulnerable to Goodhart effects due to the effectiveness of simple adversarial actions by individual webmasters, Goodhart effects would remain an aspect of concern for any technical approximation of the intended model of web search through a proxy criterion. It has been pointed out that adversarial Goodhart effects are very difficult to avoid~\cite{goodharttaxonomy}. The obvious way of keeping the proxy secret would not scale to an intelligent plurality of adversaries. The alternative, as suggested in~\cite{goodharttaxonomy}, is to choose a proxy that is conceptually {\it simple} and {\it hard to optimize} for. For example, it could be to choose a proxy such that adversaries have minimal control over the world over which optimization for it would need to function. 



\section{Link-based Web Search}

We now come to the topic of link-based web search, the predominant paradigm of need-based web search that forms the focus of our study. We outline the technology, the philosophy and extant critical observations. 


\subsection{The Technology of Link-based Web Search}

The paradigm shift in web search pioneered by Google, in the late 1990s, was to move from keyword search to predominantly link-based relevance estimation~\cite{brin1998anatomy}. Thus, it is useful to use Google's famed PageRank algorithm as typifying the paradigm. The philosophy has been spelt out by Google\footnote{https://web.archive.org/web/20111104131332/https://www.google.com/competition/howgooglesearchworks.html} as follows:

\vspace{0.05in}

\hrule 

\vspace{0.05in}

{\it PageRank works by counting the number and quality of links to a page to determine a rough estimate of how important the website is. The underlying assumption is that more important websites are likely to receive more links ...}

\vspace{0.05in}

\hrule

\vspace{0.05in}

PageRank associates every page with an {\it importance score}, one that is directly related to {\it both} the {\it number} and {\it importance} of links {\it towards} itself, often called backlinks. PageRank embeds a circularity within its formulation~\cite{franceschet2011pagerank}. For example, two pages linking to one another would have their {\it importance} scores as dependent on each other. Such circular dependencies may also appear through a third page, or a longer chain of intermediate web pages. While this engenders technical difficulties in estimation, extensive research into PageRank over the years has led to very efficient estimation algorithms~\cite{chung2014brief}. 

The design of the link-based web page importance score is fully determined by number and importance of inward links to it. This is notably, a query-agnostic estimation, independent of what query is being addressed by the web search process. Under link-based search, web pages are retrieved based on both (i) their link-based importance score, and (ii) content-based match to the query. Thus, the link-based search paradigm, while dominated by the link-based estimation of importance scores, still has a content-based matching process within it. 

This primacy of links in web search is pervasive in web search, and now pervades virtually all top search engines. For example, Bing has been observed to rely more on content relative to Google, but is still substantively dependent on leveraging links to assess page importance\footnote{https://www.forbes.com/sites/theyec/2020/07/16/how-to-rank-your-website-on-bing-in-2020/?sh=5e3940416b47}. The shift of focus from only content to predominantly links in the late 1990s has not just endured but has virtually become the only existing paradigm of web search.


\subsection{Critically Analyzing Link-based Search}

We now consider the technology of link-based search from multiple critical perspectives. 

\subsubsection{Design Choices}

If we look back at link-based search from the conceptualization in Section~\ref{ref:conceptual}, we may make some observations. First, much like the content-based search case, there is a short-circuiting of the {\it need} identification/matching step embedded within the content-based matching process. Second, the presence of the importance score in link-based search operationalizes a splitting of the overall task of {\it query/user relevance} to two parts viz., a query-agnostic link-based importance score, and a query-based content-matching score. Such splitting is technologically attractive, since they enable the application of the general-purpose filter-and-refine strategy~\cite{wood2008filter} to speed up the search process. Under the filter-refine strategy, one of the criteria could be used to filter away many web pages, so the second criterion needs to operate on a small set of web pages. In fact, there have been tremendous advances in {\it indexing} of web pages based on content (e.g.,~\cite{guo2022semantic}), to enable fast content-based filtering. In particular, it has not been observed that the dichotomous conceptualization of relevance is in accordance with any theory of human assessment of relevance, such as used in structuring automated decision making elsewhere (e.g., Gestalt theory for web page segmentation~\cite{xu2016identifying}, the psychologically motivated recency heuristic for trend prediction~\cite{gigerenzer2023psychological}). Thus, the two-part conceptualization is probably motivated by technological convenience than by anything else. 


\subsubsection{Historical Context, Philosophy and Ideology}

This link-based paradigm has been widely noted to have been motivated by the information sciences discipline of bibliometric citation analysis~\cite{lewandowski2023jasist}. Considering the number of citations received by a paper as a measure of its impact can be traced back to 1927~\cite{gross1927college}; it rests on a simple assumption that high quality scientific work will attract more citations. It has been argued that citations tend to be {\it unobtrusive}, {\it non-reactive} and notably {\it do not require the cooperation of a respondent}~\cite{smith1981citation}. To quote from~\cite{van2004new}, {\it 'When Google launched in 1997, it was said to be unspammable ...'}. One could potentially interpret these as robustness to Goodhart-like effects. However, for scientific citations, in an empirical study, it was found that non-scientific factors play a role in deciding to cite a work, even though the authors stop short of questioning the value in citation analysis~\cite{bornmann2008citation}. While citation analysis has been subject to much critique, citation analysis (and metrics based on it) is notably widely used within contemporary society to inform important decisions~\cite{impactfactor}, and we find it reasonable to assert that they are perceived in scientific circles as the most robust among available metrics of scientific achievement. Link-based importance, however, notably differs from citation analysis in one aspect in that it is not just the number of inward links that determine the importance of a web page, but also the importance of the pages that link. It has been observed~\cite{mayer2009sociometry} that this weighted formulation can be substantively traced back to sociometric work from 1953~\cite{katz1953new} where the intent was to develop a {\it status index} for social groups. It is interesting to note that there has been recent research into quantifying the importance of citations~\cite{chakraborty2016all}. 

The link-based paradigm, as pointed out critically in~\cite{carr2009big}, is a manifestation of the idea that each link represents a small bit of human intelligence, that of recommendation/endorsement (first assumption of hyperlink analysis~\cite{henzinger2001hyperlink}) and that its aggregation would create immense value. Within the broader backdrop of the politics of search engines, the focus on links creates new forms of exclusion, where web pages not having sufficient backlinks could be excluded from search engines due to not being in the crawl path or low page importance scores~\cite{introna2000shaping} (Ref: Table 1 in~\cite{introna2000shaping}). In a sharp critique of Google's PageRank~\cite{pasquinelli2009google}, Pasquinelli's 2009 work interprets PageRank as a model of cognitive capitalism. This is so since PageRank, by using links to make judgements of the value of web pages, is appropriating cognitive labor into network value, and thus functions as a global rentier of the common intellect. The operation of link-based search has also been found to be aligned with capitalism in corresponding to a form of exploitation enabled by the connexionist world~\cite{mager2012algorithmic}, dubbed to be part of the new spirit of capitalism~\cite{boltanski2005new}. These could also be read within the context of Google being observed to be facilitating exploitative digital labor~\cite{fuchs2014digital}. We have restricted our attention to critical search engine studies that particularly focus on the operations of the link-based search paradigm. 

\subsubsection{Myriad Critical Observations on Link-based Search}\label{sec:critical}

Ever since the inception of the link-based search paradigm, there have been critical observations of systemic issues. We summarize a representative sample of some such observations herein. {\it First}, the {\bf bias of web search to commercial sites} was noted, perhaps for the first time, in~\cite{van2004new} (2004), based on several computer science studies at the time. The study speculates on the possibility of link-based search being the determinant, but concludes noting that {\it 'we remain as yet unclear as to why some sites are favored over others'}. On similar lines, it was observed, in 2013, that {\it 'Google Search Favors Big Brands over Small Businesses'}\footnote{https://www.sbwebcenter.com/google-search-favors-big-brands-over-small-businesses/}. The article puts the responsibility of this behavior to PageRank construction in the context of reasons that are presented as 'natural', such as stronger backlink profiles, durable websites and better domain names. {\it Second}, {\bf myriad pernicious biases} have been noted as embedded within search results. These were brought into much popular attention through the work: {\it Algorithms of Oppression}~\cite{rahman2020algorithms} that focuses on the racism in search results. There have been several instances where major search engines, especially Google search, have been criticized for implicitly advancing {\it racist, sexist, anti-semitic and otherwise offensive values that manifest as biases}. There have also been extant observations of {\it extreme-right sources appearing in top result positions} in web search~\cite{norocel2023google}. It has been alleged that Google has often adopted a {\it whack-a-mole} approach where it reacts only when criticism is popular enough to constitute a problem for the brand~\cite{sundin2022whose}. Fairness in retrieval has evolved to an active research field~\cite{zehlike2021fairness}; yet, as often argued about technological efforts on AI ethics~\cite{katz2020artificial}, the research often focuses on using constraints and similar approaches towards ensuring diversity in the output while not being enthusiastic about identifying or addressing how unfairness comes about in the first place. {\it Third}, {\bf algorithmically embodied emissions}, as introduced in~\cite{haider2022algorithmically} outlines how algorithmic information processing systems, particularly search engines, routinely make choices that {\it 'contribute to the climate crisis and other forms of environmental destruction'.} We use an example from the authors' blog post\footnote{https://medium.com/datasociety-points/algorithmically-embodied-emissions-algorithmic-harm-and-climate-change-e2617eb4770d} to illustrate the point succinctly. A web search query {\it 'summer clothes'} could be the manifestation of multiple possible user intents, such as understanding contemporary or historical summer clothing trends, buying new summer clothes, finding clothing that produce a cooling effect, or buying pre-owned summer clothes from a second-hand shop. The authors of~\cite{haider2022algorithmically} point out that search engines consistently favor very high-carbon interpretations of search queries. In fact, an appraisal of the search results for the query may consistently point to an interpretation of {\it buying branded summer clothes}, an extremely high emission interpretation of the query. This trend, as the authors point out, is consistent across various types of queries; for example, a query {\it paris stockholm} is interpreted as a search for flight options between the cities all across the first page of search results. {\it Fourth}, there have been general observations on the {\bf reduced utility} of search engines such as Google. A recent article from a popular blog on search engines was titled {\it 'Google search is dying'}\footnote{\url{https://dkb.blog/p/google-search-is-dying}}; a relevant excerpt says {\it '... Google search results have gone to shit. You would have already noticed that the first few non-ad results are SEO optimized sites filled with affiliate links and ads'}. The article places significant emphasis on SEO actions. Another article, titled {\it 'The Tragedy of Google Search'}\footnote{\url{https://www.theatlantic.com/technology/archive/2023/09/google-search-size-usefulness-decline/675409/}} talks about Google {\it 'creaking under the weight of its enormous size'} and that {\it 'Google Search is now bloated and overmonetized'}. It is notable that these arguments often externalize the reasons for the failures; in other words, they sound like, Google or link-based search may have continued well if not for 'extrinsic' factors such as SEO actions, enormous size, and overmonetization. 

\subsection{Towards a Political Economy of Link-based Web Search}

The critical observations on link-based search, such as those noted above, often tend to be viewed as separate from one another. Well intended and valid solutions are often proposed, such as the suggestion of extending environmental risk assesments to include algorithmic harm~\cite{haider2022algorithmically}, giving more weighting to source credibility~\cite{sundin2022whose}, encouraging small businesses to invest in SEO to compete better with big ones\footnote{https://www.sbwebcenter.com/google-search-favors-big-brands-over-small-businesses/}, exhorting users to specialize queries to include source names (e.g., using the {\it site:} operator\footnote{https://dkb.blog/p/google-search-is-dying}) and promoting algorithmic capacity for search result diversification\footnote{https://fair-trec.github.io/}. The above strategies differ on the extent to which they are technocratic, qualitative or quantitative. These also reflect in the potential impact these could have. While they are all important in their own ways, we take a complementary approach and posit that it would be beneficial to develop an understanding of how the myriad issues of web search are {\it systemic issues} that have their roots in the {\it totality of the socio-technical ecosystem and social relations embedded in the paradigm of link-based web search}. This would help understand the core phenomena of which the observed issues are myriad manifestations. This requires developing an understanding of the political economy of link-based search. The nature of the design choices, historical context, philosophy and ideology, all play vital roles in understanding the political economy of link-based search. 

\begin{figure*}[h]
\begin{center}
\includegraphics[width=0.7\textwidth]{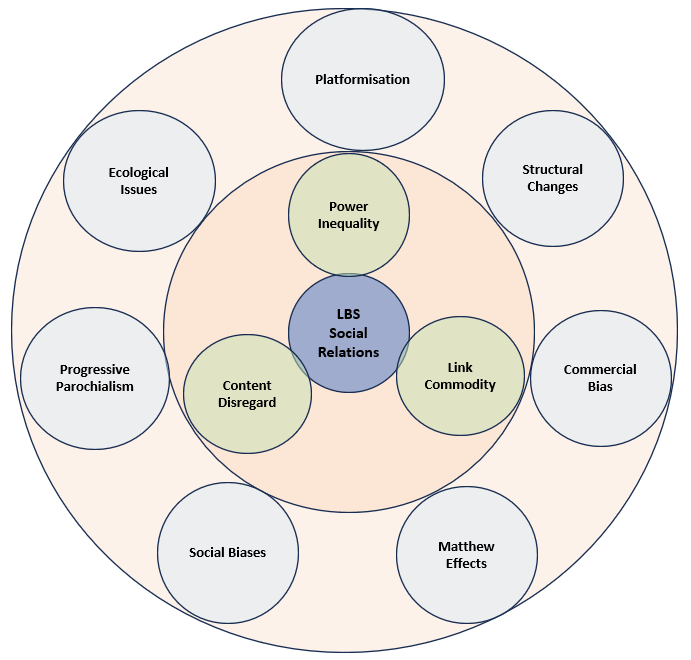}
\caption{Illustrative Summary of our Theoretical Framework}
\label{fig:summary}
\end{center}
\end{figure*}

\section{The Political Economy of Link-based Web Search}

We now delve into a few steps towards the aforementioned ambitious goal of developing an understanding of the political economy of link-based web search. We take a consequence-focused approach and outline our conceptualization of the political economy of link-based search in Fig~\ref{fig:summary}. We posit that the drawbacks of link-based search can be traced back to its socio-technical ecosystem; thus, the social relations of link-based search (abbreviated LBS) is at the centre of our theoretical framework. We partition the consequences of LBS social relations into {\it first order effects} (inner circle in Fig~\ref{fig:summary}) and {\it second order effects} (outer circle). In what follows, we present how we can trace the pathways from LBS social relations to the first order effects, and consequently how the second order effects could be seen as emanating from LBS social relations and the first order effects. We do note herein that there are synergistic, complementary and other kinds of relations across the various consequences we consider in our theoretical framework; thus, our structuring of consequences as first and second order effects are to be seen as an effort towards narrative clarity rather than one towards a holistic representation. 

\subsection{First Order Consequences}

We now consider the first order consequences in our theoretical framework and their interrelationships. 

\subsubsection{Power Inequality} We observe that there are two ways in which LBS social relations engender power inequality across web pages (and thus, webmasters). {\it First}, The link-based formulation of a query/need-agnostic importance score is deeply embedded within LBS, as observed earlier. This creates an {\it a priori} gradient in importance across web pages, one that acts in substantive ways as a determinant of eventual visibility through search engines. For example, across two web pages, the higher importance of one web page can offset any deficit in content-based relevance relative to the other, and thus, the former has a higher propensity to achieve a higher rank. In other words, a web page with a lower query/need-relevance than another may be ranked higher than the latter owing to its higher importance score. Under the ethos of relevance-based search (Ref Sec~\ref{ref:conceptual}), this could be seen as a violation of equality of opportunity. Similar phenomena are observed in social media where engagement metrics such as the number of likes function as an a priori importance measure~\cite{vaidhyanathan2018antisocial}; high engagement posts, while potentially crowding out relevant posts for the user, are aligned with the interest in maximizing user attention. {\it Second}, given the formulation of the importance score as an importance weighted aggregate of links, a web page with a high importance score is much better positioned to bestow another with an improvement in importance score through handing out a link. This enables actions of webmasters of high-importance pages to have higher say in the operation of the web search process. Thus, web pages with higher importance scores wield much higher power within the socio-technical ecosystem of LBS. 

\subsubsection{Link Commodity} As observed earlier in content-based search, the presence of web search as an arbiter of user attention creates incentives for webmasters to undertake actions to enhance their visibility. However, in contrast to content-based search where these incentives lead to individual-level content changes, the dominance of links in LBS requires that pathways to enhance inward links be sought. Unlike content on the web page, inward links to the web page from important\footnote{Important in accordance to the importance score.} web pages {\it cannot} be created by the webmaster of the page in question. Instead, they ought to be created by the webmasters of important web pages. Consequently, towards improving visibility, the webmaster of the page in question would need to {\it publicize} their web page, in perhaps a targeted way to gather eyeballs from webmasters of 'important' web pages, in order to ensure some inward links to their page, and thus, a high-enough importance score. Given the prime importance of this activity called {\it link building} in ensuring visibility, guides for link building (such as those from Ahrefs\footnote{\url{https://ahrefs.com/blog/link-building/}}) suggest at least two main options for webmasters: (i) {\it ask for a link}, or (ii) {\it purchase a link}. The first option aligns with {\it marketing}, a tool within a market society, of which, advertising is a popular form. The second option of purchasing links implies a {\it direct commodification} of links within the visibility oriented market created by link-based web search. While the trading of link commodities between webmasters may not be very visible to end users, contemporary pervasive practices such as offering discounts for reviews are a manifestation of the commodification of links\footnote{https://www.fera.ai/blog/posts/offering-incentives-for-reviews-can-i-offer-a-discount-or-loyalty-points-for-leaving-a-review}; review links are said to account significantly for {\it local search} in Google\footnote{https://moz.com/learn/seo/local-ranking-factors}. The emergence of the link commodity is a substantial consequence of LBS social relations. 

\subsubsection{Content Disregard} With links and link-based importance scores becoming a dominant determinant of visibility, the role of content is crowded out from web search. Webmasters increasingly invest time and energy on link building as against creating useful content, running contrary to the conceptualizations of web search discussed in Sec~\ref{ref:conceptual}. That LBS cares much more about the {\it structure or form} than the {\it content} of websites aligns it with capital which has been studied for a general disregard of content, or the subordination of content to form. Marx points out that capital is only concerned with specific use-values (i.e. specific content) insofar as they have the general form of a commodity, or in other words, that they have an exchange-value (cf. use value) and can be sold (Capital Vol 1~\cite{marx1867capital}, p461). Link-based web search similarly focuses on the formal dimension of link metrics, rather than the specific content of websites. 


\subsection{Second Order Consequences}

We now consider the second order consequences, their interrelationships as well as their relation with first order ones. 

\subsubsection{Platformization} Search engines are increasingly being viewed as a primary source of information, a function traditionally aligned with media. If so, what kind of a media are search engines? Traditional media, such as newspapers, operate as a two-sided market~\cite{rochet2003platform}, with readers and advertisers as two sides. The content is curated and owned by the platform. It has been observed that search engines~\cite{van2017political}, on the other hand, operationalize a separation of content from delivery, with content being provided by websites outside the control of the search engine. This makes search engines a kind of multi-sided platformized media. This platformization creates a new dimension of reciprocal {\it indirect network effects} between users and webmasters where the proliferation of websites creates more user demand and vice versa. This is illustrated schematically in Fig 2 within~\cite{van2017political}. We observe that link-based search {\it adds another dimension} to such network effects, and consequently to enhanced platformization, with links being a manifestation of network value; here, we stick with the founding assumption of link-based search of considering link as a proxy of importance/value. To put it simplistically, proliferation of links among existing websites are capable of catalyzing user demand, and vice versa. Thus, the emergence of link-based search enhances platformization of search engines in substantive ways. Platforms need to be viewed as not just a technological infrastructure, but as a political enabler/catalyst of a multi-sided market~\cite{gillespie2010politics}. When viewed from the economic theory of conceptualizing platforms as markets~\cite{rochet2003platform}, link-based search, through deepening platformization, enhances the marketization of search. Observations such as overmonetization (Ref: Sec~\ref{sec:critical}) could then be read as fallouts of such consequences. 

\subsubsection{Structural Changes} It was observed earlier that inward links functioning as a determinant of visibility leads to a commodification of links. This facilitates {\it inorganic creation} of links through trade in the marketplace of links. Notably, this pathway of inorganic creation of links adds to the pace of overall creation of links. This could potentially make link creation a more dominant factor in the evolution of the web, as compared to other factors such as creation of content/pages. It is not obvious to us as to what kind of large-scale effects this could precipitate. Yet, studies point to significant structural changes in the web in the first two decades of this millenium. In 2000, a pioneering work~\cite{broder2000graph} on analyzing the web structure points out the existence of a single large set of web pages connected through hyperlinks (called a {\it connected component} in graph terms) and several small components disconnected from the large set; this structure, called the {\it bow-tie} structure, is not without detractors~\cite{meusel2014graph}. Two decades later, an analysis of web structure finds a sharp departure from the bow-tie structure, towards multiple local bow-tie structures~\cite{fujita2019local}. While it is not clear as to what effect, if any, the emergence of the link commodity may have had on web structures at a macro level, the evolution does suggest the emergence of stronger localized traffic flows. Notably, the enhanced connectivity is to the benefit of {\it navigational media}~\cite{van2011navigational}, the new ethos of web media which relies on the conceptualization of traffic distribution as the source of revenue. In other words, it strengthens the hands of new models that rely on traffic as opposed to traditional models that rely on audiences as the source of revenue (e.g., traffic commodity~\cite{van2008history}). This could be catalyzing the incorporation of navigation-like elements in traditional media, such as QR codes in print media~\cite{nath2020factors}. 

\subsubsection{Commercial Bias} The correlation of being linked and being visible, and the emergence of the link commodity, makes web page visibility directly correlated with their webmasters' monetary spending for link building. Profit-focused content creators who look to translate enhanced visibility to profits are implicitly best positioned to spend money on link building (through express spending or offering discounts in exchange for links), since they can make the enhanced web visibility translate to monetary outcomes. Within this market-based dynamics of link-based search, an academic content creator who may create quality content is ill-positioned to achieve visibility in the link-based era (cf. content-based era), since they aren't looking at web visibility as a source of revenue, and thus can't bank on enhanced visibility to offset monetary costs incurred in achieving it (and, are not well poised to undertake monetary spending for link building). This makes commercially oriented web pages more linked within the web, bringing about a commercial bias in web search. This makes link-based search implicitly favorable to markets, and thus, to the interests of capital, slowly transforming the Web from an information system to a marketplace. This could be seen as a particularly narrow and latent manifestation of adversarial Goodhart effects, where link-based search produces an ecosystem that is conducive for a narrow spectrum of adversaries, those oriented commercially, to appropriate value from web search through link building. 

\subsubsection{Social Biases} We have observed that link-based search moved the heavy lifting in web search from content-matching to link-based scoring. What would this mean for the visibility of social biases within web search? In a somewhat tangential domain, that of social media, it has been found that {\it likes} on social media can convey myriad personality traits and political views, among others~\cite{youyou2015computer}. It has been pointed out that likes often convey more than a user would themselves like to\footnote{https://ideapod.com/facebook-likes-expose-think-according-computer-scientists/}. Against this backdrop, it may be reasonable to speculate that links could convey a lot more latent traits than what is willingly offered through the contents of web pages. Racism is of several types; of which {\it inadvertent}, {\it habitual} and {\it explicit} are examples~\cite{facesofracism}. For example, while the content of web pages may convey {\it explicit} racism, {\it inadvertent} and {\it habitual} racism may not appear in the content but could manifest through links. Thus, link-based search, in contrast to content-based search, could be interpreted as offering new channels of flow of social biases from society to web search results. It has been observed that actual web traffic patterns are more racialized than the hyperlink networks~\cite{mcilwain2017racial}; this may be read in the light of pervasive usage of search engines for user navigation, making traffic patterns directly dependent on the search engines (and thus, link-based importance scoring). Further, extant studies explore how reliance on source popularity leads to a far-right politics of exclusion~\cite{norocel2023google}; it is notable that popularity could be correlated with link-based importance scores. Thus, link-based search could be, at least partially, responsible for the observations of widespread social biases in search results. 

\subsubsection{Matthew Effects and Progressive Parochialism} The Matthew effect~\cite{rigney2010matthew} is most commonly described using the adage-like statement: {\it the rich get richer and the poor get poorer}. However, it is best described as one of {\it preferential attachment}, where rewards get distributed based on current possessions\footnote{Thus the poor needn't get poorer in an absolute sense, only in a relative sense to the rich}. The social relations of link-based search are particularly oriented towards the Matthew effect. As an example, a web page (actually, the webmaster behind the web page) that is rich with links commands a higher visibility which further enhances their ability to acquire more links. This can be contrasted with the higher visibility of high use value web pages in content-based search, which does not produce a feedback loop; that is, visibility doesn't directly or automatically lead to alteration of content to further enhance visibility. The Matthew effect arguably accentuates every consequence of link-based search. For example, the initial commercial bias could progressively lead to steeper skews leading to a big business bias, and an eventual facilitator of monopolization. Towards understanding this, consider a web-based food delivery platform which is very popular, and thus are bestowed with high {\it page importance} scores. They are able to translate their high {\it page importance} score to rope in more new businesses (e.g., takeaways) into their platform since their links will ensure higher visibility to the businesses than a link from a less popular food delivery platform. These all lead to a progressive parochialism in the operation of link-based search. 


\subsubsection{Ecological Issues} Here, we consider the world of ideas underpinning ecosocialism~\cite{huan2010eco}. The foundational urge of capital to maximize wealth/value leads to a propensity towards {\it commodification of everything}~\cite{harvey2007brief} so that their values can be realized within markets to enlarge profits. Such extreme commodification spares little, and the natural environment is not immune to the urge. Indeed, as Marx points out, this leads to an {\it 'irrepairable rift in the interdependent process of social metabolism, a metabolism prescribed by the natural laws of life itself'}~\cite{marx1894capital}. This has been described as {\it 'metabolic rift'}~\cite{foster2000marx}, a breakdown of the metabolic relationship between humans and nature. In other words, advanced forms of capital - embedded in big businesses - predicate an unmistakable anti-nature ethos. Against this backdrop, we may reconsider the various observations made so far. The commodification of links enhances the market ethos on the Web, and the social relations of link-based search engender a commercial bias. The predominance of commercial bias and big business bias in web search translates, somewhat directly, to the privileging of commercially oriented, and thus ecologically unsustainable, interpretations of search queries, as observed in~\cite{haider2022algorithmically,haider2023google}. Thus, ecological unsustainability may be seen as being a fallout of the social relations encoded within the socio-technical ecosystem of link-based search, and intricately linked to commercial biases. 

\subsection{Allied Noteworthy Aspects}


We now consider some broader and allied aspects herein. 

\subsubsection{Deterioration of Utility} The discussion so far considered myriad consequences of the social relations and socio-technical ecosystem of link-based search. They all converge on one broad trend, that of catalyzing a deterioration of the utility of link-based search as a search paradigm. While link-oriented platformization promotes the marketization of search, it could be fragmenting the web into silos. The myriad social biases and commercial biases could be racializing and commercializing web search, taking web search away from being able to effectively mediate between queries and web pages. Web search could be privileging unsustainable interpretations, aggravating the ecological crisis. The accentuation of the above phenomena through Matthew effects could be creating a subtle but virulent feedback loop parochializing web search. The complementarities, overlaps and synergies among the factors could eventually trigger significant drops in the utility of the link-based search as a paradigm of information seeking on the Web. Under the light of these observations, we could re-read allegations of futility of link-based web search, as observed in Sec~\ref{sec:critical}. 

\subsubsection{The Mode of Production in Link-based Web Search} We now attempt to situate link-based search within the concept of {\it automation}, an important aspect of critical political economy scholarship. Automation, within traditional manufacturing sectors, is often described as a change in the {\it organic composition}~\cite{shaikh1990organic} of the mode of production, that which increases the proportion of machinery to labor in the production process. Marxist economics sees automation as an intrinsic tendency of capital~\cite{rowthorn1985organic}, which works in lockstep with its urge to enhance profits~\cite{marx2005grundrisse}. Machinery is itself produced using manual labor in the past, and thus, often described as {\it dead labor}~\cite{marx1867capital}, i.e., an ossified manifestation of living labor from the past. We posit that link-based search sits in a somewhat uneasy way within the broader understanding of automation.

Link-based search, as observed earlier, restructures the retrieval process into an importance computation step and content-based matching. Let us consider the {\it nature of work} involved in these separate 'production' processes. The page importance computation process has a very simple technical form, but relies heavily on the existence of an up-to-date {\it link infrastructure} within the Web. The link infrastructure, as we have seen before, is the product of webmaster actions, and thus, may be described as ossified historical and globally distributed labor. The matching process, on the other hand, is the query-dependent process and could be potentially complex, given the need to ensure that web pages that satisfy the user's need is ranked sufficiently highly. This is labor performed by matching algorithms, which are themselves product of ossified labor by programmers. Under this lens, the importance computation and matching process are both heavily aided by ossified labor, the first by labor ossified primarily within the {\it link infrastructure} and the second by labor ossified within {\it matching algorithms}. 


A crucial difference between the above two forms of ossified labor is notable. The {\it link infrastructure} is labor by webmasters who are not paid by the search engines. By creating each link, the webmaster enriches the link infrastructure through undertaking latent and unpaid digital labor. On the other hand, the {\it matching algorithms} are produced by digital labor by programmers and software architects within the search engine company, and is thus paid digital labor. Thus, the shift from content-based search to link-based search changes the organic composition of web search from being dominated by the fruits of paid digital labor to one being increasingly the product of unpaid digital labor. This is heavily aligned with the profit urge of capital since unpaid digital labor is intrinsically more profit oriented than paid labor. 


Further, it is notable that the quality of web search relies heavily on the link infrastructure being kept up to date, to serve the changing needs of web search users. On the other hand, matching algorithms may only need to be updated much less frequently. Thus, in the long run, the maintenance of the link infrastructure could be viewed as having characteristics of living labor. Thus, the shift from content-based to link-based search may be seen as a shift towards living labor. Given our focus on link-based search, we keep the labor of web page creation outside our remit.

\section{Conclusions}

We considered the political economy of the paradigm of link-based web search, and proposed a theoretical framework towards enhancing our understanding. Based on our analyses, we find it reasonable to assert that the emergence of link-based search is favorable for the entrenchment and accentuation of libertarian ethos on the Web. The socio-technical environment posited by link-based search appears highly correlated with the interests of capital. From the vantage point of our theoretical framework, several extant critical observations of link-based search may be seen as predicated by the social relations embedded in link-based search. We also observed how such myriad effects synergize in a deterioration of utility of link-based search, and could potentially point to an eventual exhaustion of the utility of link-based search. Our analyses also suggest that pathways to address the separate consequences should not foreclose a search for alternatives to the libertarian core of link-based search. In future work, we intend to look at how the link-based ethos has pervaded most of our digital lives through facets such as social media, information aggregation platforms and the platform-oriented gig economy. 



\bibliographystyle{ACM-Reference-Format}
\bibliography{refs}

\end{document}